\documentclass{article}

\usepackage{arxiv}

\usepackage[utf8]{inputenc} 
\usepackage[T1]{fontenc}    
\usepackage{hyperref}       
\usepackage{url}            
\usepackage{booktabs}       
\usepackage{amsfonts}       
\usepackage{nicefrac}       
\usepackage{microtype}      
\usepackage{lipsum}		
\usepackage{amssymb}
\usepackage{amsmath}
\makeatletter
\newif\if@restonecol

\usepackage[linesnumbered,ruled,vlined]{algorithm2e}
\usepackage{algpseudocode}
\usepackage{graphicx}
\usepackage{color}

\def\onedot{{.}}
\def\etal{{et al}\onedot{ }}
\title{Secure Software-Defined Networking Based on Blockchain}

\author{
  Jiasi~Weng\\
  College of Information Science and Technology\\
  Jinan University\\
  \texttt{wengjiasi@gmail.com} \\
   \And
 Jian~Weng \\
  College of Information Science and Technology\\
  Jinan University\\
  \texttt{cryptjweng@gmail.com} \\
  \And
 Jia-Nan~Liu \\
  College of Information Science and Technology\\
  Jinan University\\
  \texttt{j.n.liu@foxmail.com} \\
  \And
 Yue~Zhang \\
  College of Information Science and Technology\\
  Jinan University\\
  \texttt{zyueinfosec@gmail.com} \\
}

\begin{document}
\maketitle

\begin{abstract}
Software-Defined Networking (SDN) separates network control plane and dat a plane, which provides
a network-wide view with centralized control (in the control plane) and programmable network configuration for data plane injected by SDN applications (in the application plane).
With these features, a number of drawbacks of the traditional network architectures such as static configuration, non-scalability and low efficiency can be effectively avoided.
However, SDN also brings with it some new security challenges, such as single-point failure of the control plane, malicious flows from applications, exposed network-wide resources and a vulnerable channel between the control plane and the data plane.
In this paper, we design a monolithic security mechanism for SDN based on Blockchain.
Our mechanism decentralizes the control plane to overcome the single-point failure while maintaining a network-wide view.
The mechanism also guarantees the authenticity, traceability, and accountability of application flows, and hence secures the programmable configuration.
Moreover, the mechanism provides a fine-grained access control of network-wide resources and a secure controller-switch channel to further protect resources and communication in SDN.
\end{abstract}


\section{Introduction}\label{sec:introduction}
\subsection{Motivation}
In recent years, Software-Defined Networking (SDN) has received great attention from both the research community and the industry. For example,
Google has already implemented an SDN architecture, Google B4, on its data centers \cite{jain2013b4}.
By separating the network control and the data plane, SDN overcomes several limitations of the traditional networks such as static configuration, non-scalability and low efficiency.
Due to the logical centralization of the control plane and the programmability of the configuration for the data plane, SDN provides a global view of network resources that enhances the performance and flexibility of the underlying networks.
\begin{figure}[!t]
\centering
\includegraphics[width=5in]{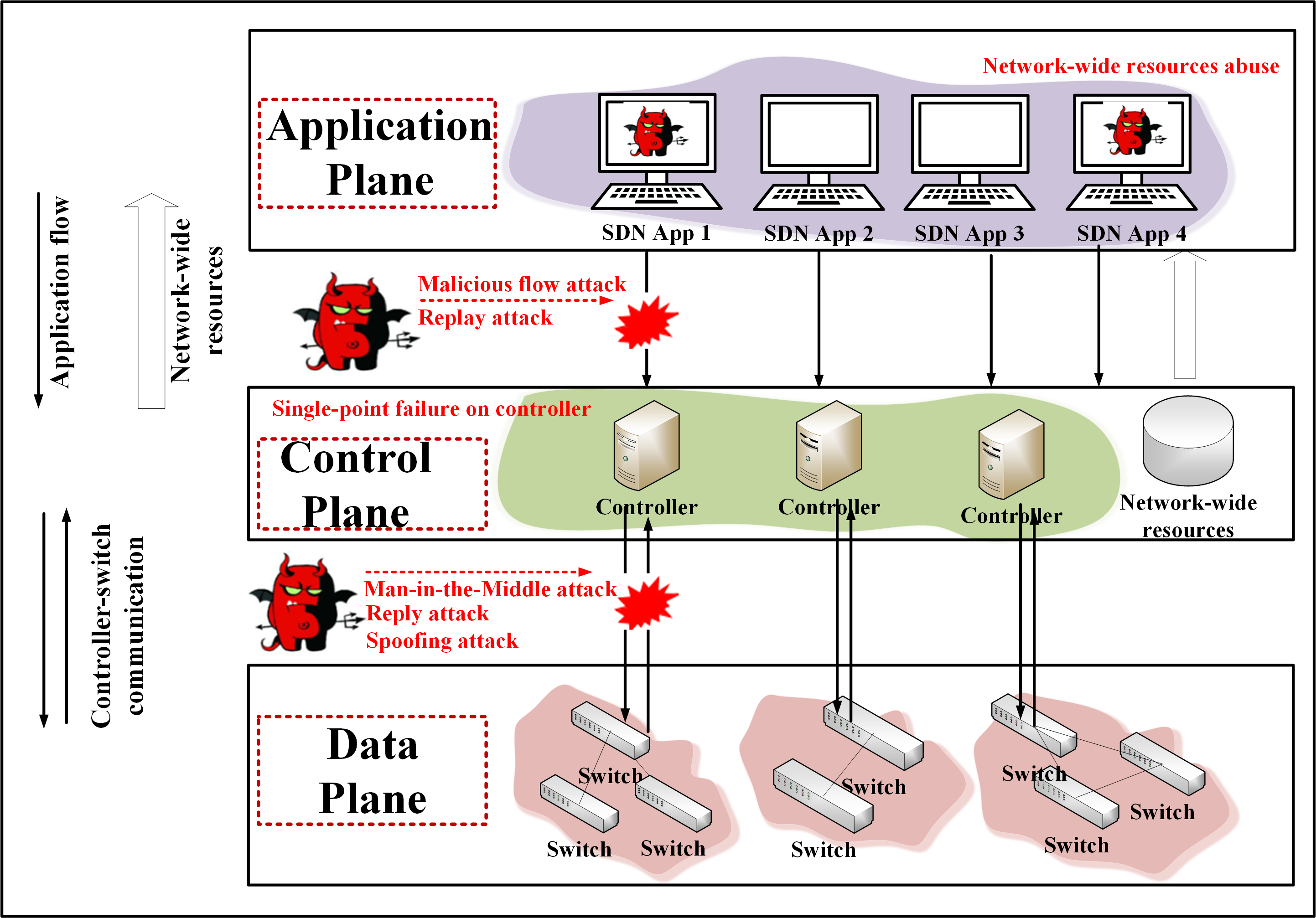}

\caption{The overview of attacks on SDN architecture}
\label{fig:attacks}
\end{figure}

However, applying SDN into networks also introduces some new security issues, which has been stressed in recent years among SDN researchers \cite{pisharody2017brew, zhang2017secure, chung2013nice, bonola2015streamon}.
The overview of attacks on SDN architecture is shown as Fig. \ref{fig:attacks}.
Most of these security issues exist in the application plane and the control plane.
On one hand, without authentication, applications may inject malicious configurations into network devices at will, which could reduce network availability, reliability and even lead to a network breakdown. Note that application flows are network configurations sent by applications and are managed by controllers who install network configurations into switches. Loss of traceability and accountability of application flows  may cause trouble for network debugging.
In SDN, tracing and auditing application flows and network states can help monitor and replay network states or debug a broken-down network, and network behaviour information also can be applied to recognize attack patterns \cite{Kim2004A}.
On the other hand, the control plane also exposes network-wide resources to all applications that will open a door for malicious applications. 
Moreover, adopting a single controller may result in a single-point failure which can become an attractive target for DoS attacks.

Finally, the lack of authenticated controller-switch communication channel could lead to more severe security threats when the configuration-complex TLS protocol (Transport Plane Security, TLS) is not adopted. Adversaries can launch man-in-the-middle attack and eavesdropping attack by seizing all packages between controllers and switches \cite{benton2013openflow, cui2016fingerprinting}.
In addition, malicious switches can also launch spoofing attacks by faking identities (e.g., IP addresses) which may lead to DoS/DDoS attacks \cite{wu2016low, merlo2014denial}.
In order to address the above security problems, many proposals have been made in recent years                                                                        \cite{Tantar2014Cognition,zaalouk2014orchsec,porras2012security,hinrichs2008expressing,son2013model,khurshid2012veriflow,handigol2012debugger,ballard2010extensible,
tootoonchian2010hyperflow,koponen2010onix,phemius2014disco,monaco2013applying,ferguson2013participatory,
matsumoto2014fleet,Wen2013Towards,Nayak2009Resonance,Peng2009Improved,Santos2014Decentralizing}.
These proposals each targets at a specific security issue, but to the best of our knowledge, no attemps have been made which address all the common security issues simultaneously with a monolithic architecture. Specifically, existing works that provide application flow authentication or flow secure constraint are to extend an individual secure module on a controller rather than a monolithic secure module in a multi-controllers environment. Additionally, existing role-based access control schemes on network-wide resources are not fine-grained enough. Moreover, optional TLS protocol and other cryptography-based authentication protocols presented in \cite{Peng2009Improved,Santos2014Decentralizing} require multiple interactions (also called multiple pass) to build a controller-switch communication channel. In a network with the physically decentralized control plane, simply combing those existing schemes fails to effectively solve all the common security issues simultaneously, because all secure modules must work seamlessly among multiple controllers.
\subsection{Contributions}
In the paper, we present a Blockchain-based monolithic secure mechanism to effectively address multiple common security issues in SDNs.
In particular, the paper makes the following contributions.
%

$\bullet$ Our mechanism decentralizes the control plane into multiple controllers while maintaining consensus among all controllers on network-wide resources.

$\bullet$ To overcome the weakness of lacking traceability and accountability of application flows, all flows and network behaviours are recorded on Blockchain so that the network states can be easily replayed for auditing and debugging.

$\bullet$ By assembling a lightweight and practical Attribute-Based Encryption (ABE) scheme, the access permissions of each application on network resources are defined and enforced to avoid resource abuse.

$\bullet$ The effective authentication protocol HMQV (one-pass) is combined with Blockchain to protect the communication channel between the controller and switch against active attacks.
%
\section{Related Work}
A variety of security measures directed against various security threats among different planes of SDN architecture have been proposed. In terms of application plane, FRESCO\cite{Shin2013FRESCO} is a security development framework compatible with OpenFlow for SDN applications. Cognition\cite{Tantar2014Cognition} was proposed to enforce the security of applications by defining cognitive functions. OrchSec\cite{zaalouk2014orchsec}, an architecture considering the advantages of network-visibility and centralized control provided by SDN, was introduced to develop security applications. FortNOX\cite{porras2012security} extended NOX controller to provide security constraints on flow rules and an role-based authentication scheme for SDN applications. FSL\cite{hinrichs2008expressing} presented a security authentication framework for flow-based network policies. Similarly, Son \etal  proposed Flover\cite{son2013model} and Khurshid \etal  presented VeriFlow\cite{khurshid2012veriflow} to verify dynamic flow policies. With the requirements to audit and track network process, Hadigol \etal\cite{handigol2012debugger} studied network event debugger enabling network manager to track the root cause of a network bug. OpenSAFE\cite{ballard2010extensible} was proposed to support security auditing and Flow Examination to analyze network traffic and filter network package. On the control plane, a lot of frameworks with a decentralized control plane for OpenFlow were presented. HyperFlow\cite{tootoonchian2010hyperflow} was built on a distributed file system to realize network event distribution among multiple controllers. Onix\cite{koponen2010onix} implemented a physically distributed but logically centralized control platform to avoid threats brought from a single controller. Also, SDN control frameworks such as ONOS\cite{krishnaswamy2013onos}, DISCO\cite{phemius2014disco}, yanc \cite{monaco2013applying}, PANE\cite{ferguson2013participatory}, and Flee\cite{matsumoto2014fleet} supporting the distributed network logic. In order to secure network-wide resources, some security schemes \cite{Wen2013Towards, Nayak2009Resonance}were devoted to provide access control mechanism to protect resources from unconcerned SDN applications. As for controller-switch channel, Transport Plane Security (TLS) was adopted on OpenFlow specification. However, it became optional due to the insufferable drawbacks of TLS. Apart from authenticated controller-switches communication, those were excellent security measures and systems, such as FRESCO \cite{DBLP2013FRESCO}, FloodGuard \cite{WangXG15FloodGuard}, AVANT-GUARD \cite{ShinYPG13AVANT-GUARD}, FLOWGUARD \cite{Hu2014FLOWGUARD}, SE-Floodlight\cite{floodlightcontroller}, SoftFirewall\cite{koerner2014oftables}, CPRecovery \cite{suh2014building} and so on \cite{ShinXHG16}.
\section{Organization}
The rest of the paper is organized as follows. In Section \ref{sec:preliminaries}, we provide a quick overview on Blockchain, Attribute-Based Encryption, and the HOMQV protocol. Then, we describe
our security requirements on OpenFlow/SDN in Section \ref{sec:secRequire}. In Section \ref{sec:design}, we present the design of the Blockchain-based monolithic module. We analyze security issues of the construction in Section \ref{sec:analyze} and in the Section \ref{sec:implement} we present a prototype implementation of our mechanism. Lastly, we conclude the paper in Section \ref{sec:conclusion}.
\section{Preliminaries}\label{sec:preliminaries}
In this section, we give a brief introduction about Blockchain, Attribute-based encryption and HOMQV protocol.

\textbf{Blockchain} is originated from bitcoin and becomes an emerging technology as a decentralized, sharing, immutable database \cite{nakamoto2008bitcoin, croman2016scaling, kogias2016enhancing, aitzhan2016security, swan2015blockchain, cota2017racoon++}.
Data in Blockchain is stored into blocks which are maintained as a chain.
Each block of Blockchain contains a timestamp and the reference, i.e., the hash of a previous block.
Blockchain is maintained in a peer-to-peer network.
The majority of Blockchain network nodes run a consensus protocol to achieve an agreement to generate a new block.
Meanwhile, the data, also called \emph{transactions}, in the new block are also confirmed due to its consensus protocol.
Consensus protocols in the Blockchain setting can be implemented by several different agreement methods, such as POW-based (Proof of Work), BFT-based (Byzantine fault-tolerant) and POS-based (Proof of Stake).
We call them Blockchain protocols.
In the paper, we focus on the BFT-based Blockchain protocol \cite{duan2014hbft}.
This kind of protocol promises instant consensus\cite{lamport1982byzantine}.
In the meantime, Blockchain based on the kind of protocol can balance scalability and performance well, among which scalability means the number of participants and performance includes throughput and latency.
Vukolic \etal \cite{vukolic2015quest} showed BFT-based Blockchains behave excellently in network performance and guarantee instant consensus.
It also presented that BFT-based Blockchains possess the excellent ability to support the large capability of clients.
By applying BFT-based Blockchains to our module, controllers in SDN play the roles as clients of the Blockchain, which demonstrates excellent network scalability of controllers.
On the other hand, there are two ways to write data on the Blockchain: \emph{transaction} and \emph{smart contract}.
Smart contract is a program that can automatically execute the partial and total operations pre-defined in the contract and output values as evidences supporting to be verified on Blockchain.
Smart contract usually provides an outer interactive interface and the interaction can be verified based on the cryptography so that smart contract is executed in strict accordance with the predefined logic.
In our context, we build security protocols by smart contract that can be automatically executed when the predefined conditions are triggered on the Blockchain.
Owing to the immutable recorded transactions and results of executed protocols, Blockchain supports network to trace any record linked with a specific time point.
Eventually, Blockchain enables the network-wide data to share some valuable features such as reliability, non-repudiation, traceability and auditability which are adapted to our security requirements.
Thus, this paper is interested in BFT-based Blockchains adapted to our security goals to construct a Blockchain-based secure module which strengthens the security in SDNs.

\textbf{Attribute-based encryption} (ABE) contributes to a fine-grained access control for encrypted data. In an ABE system, a kind of encrypted resources are labeled with a set of descriptive attributes and a specific access structure which are associated with a private key of an access-user. It determines the kind of encrypted resources that can be decrypted by the access-user with the private key satisfying the pre-defined access structure\cite{goyal2006attribute,bethencourt2007ciphertext, yao2015lightweight, guo2014cp, liu2015traceable, jung2015control}.
We employ a lightweight ABE scheme which possesses execution efficiency and low communication costs\cite{yao2015lightweight} to achieve our secure and efficient requirements for access control. As for the framework of ABE, four algorithms are introduced as follows and they will be used as black boxes.
\begin{itemize}
\item $(PK,MK)$${\sf =Setup}(\kappa)$: Taking input of the security parameter $\kappa$, this algorithm outputs the public parameters $PK$ and a master key $MK$.

\item $E$${\sf=Encryption}(m, attr, PK)$: This is a randomized algorithm that takes as input a message $m$, a set of attributes $attr$, and the public parameters $PK$. It outputs the ciphertext $E$.

\item $D$${\sf =KeyGeneration}(\textbf{A}, MK, PK)$: This is a randomized algorithm that takes as input an access structure $\textbf{A}$, the master key $MK$ and the public parameters $PK$. It outputs a decryption key $D$.

\item $M$${\sf =Decryption}(E,D)$: Taking as input the ciphertext $E$ that was encrypted under the set $attr$ of attributes, the decryption key $D$ for access control structure $\textbf{A}$ and the public parameters $PK$, it outputs the message $M$ if $attr \in \textbf{A}$.
\end{itemize}

\textbf{One-pass HMQV  protocol (HOMQV)} is a high-performance securely authenticated protocol which combines security, flexibility and efficiency \cite{krawczyk2005hmqv}. Its security has been proved in \cite{halevi2011one}.
Specifically, it uses a cyclic group $G$ of prime order $q$ and is generated by a given generator $g$. In the initial step, there are two communication parties Alice ($ID_{Alice}$) and Bob ($ID_{Bob}$) with the long-term keys $A=g^a$, $B=g^b$, respectively. Before a key-exchange protocol, Bob first checks the key $A$ sent by Alice that $A\in G'$ (if not it aborts). Then Bob randomly chooses $y\in_{R} Z_q$, computes $Y=g^y$ and sends it to Alice. Bob also computes a session key $H(\sigma,ID_{Alice},ID_{Bob},Y)$ where $\sigma=A^{(y+eb)}$ and $e=H'(Y,ID_{Alice})$. When receiving $Y$ and $ID_{Bob}$, Alice checks $Y$ and Bob's public key in $G'$ (if not it aborts) and then computes the session key $H(\sigma',ID_{Bob},ID_{Alice},Y)$ where $\sigma'=(YB^e)^a$. Finally, Alice and Bob share the same session key because of $\sigma=\sigma'$ and the triple $(ID_{Alice},ID_{Bob},Y)$ is regarded as the session id.
However, applying the basic HOMQV protocol to controller-switch communication, controllers as receivers fail to resist reply attack since the protocol only has one pass. In addition, it needs a certificate authority to update the long-term keys of two parties in the basic protocol. Fortunately, the two security issues can be overcome by using the Blockchain, which enables a security-strengthen protocol applied into the authenticated controller-switch communication.
\section{Security Requirements}\label{sec:secRequire}
In this section, we list security requirements which should be achieved.

\textbf{Application flows authentication.} The nature of high programmability and configurability on network devices of OpenFlow/SDN forces us to pay more security attention on the application plane.
Applications (e.g., traffic engineering) provide a variety of management services for network by configuring application flows versus new threats for network.
For example, a malicious or compromised application may inject malicious application flows into network devices thereby leading to a dramatic consequence.
Therefore, to authenticate application flows from legitimate applications and verify the authenticity of application flows are significant for the configurable OpenFlow/SDN.

\textbf{Application flows tracing and accounting.} Traceability and accountability for application flows can assist operators to troubleshoot network once a network device breaks down or suffers from abnormal network behaviors.
On the other hand, for the scenarios of flow arbitration, a flow arbitration system with the duty to arbitrate conflict flows, needs to identify which flows are sent by which applications\cite{porras2015securing} where traceability and accountability for application flows are urgent to be provided.

\textbf{Network behaviours auditing.} An audit system provides periodic auditing for network behaviours (network events associated with resulting network states), which helps to enforce the stability and robust the security of OpenFlow/SDN-based network.
Analyzing from auditing results by linking network events with respective network states in current time, operators can make adjustment of the network management and enhance network performance next time.
Furthermore, relying on the audit system, attack pattern recognition also can be supported to resist future network attacks.

\textbf{Secure access control on network-wide resources.} It faces some potential threats that network-wide resources on the control plane are exposed to all applications.
For instance, Hartman \etal [35] presented a kind of network security applications serving for firewall or intrusion detection that can access network resources of the firewall.
A malicious application may abuse the resources by utilizing the instance of the kind application to bypass the firewall.
Consequently, it is necessary to construct a secure access control mechanism which is customized to applications according to their categories and the network scope they are supposed to contribute to.

\textbf{Decentralization of control plane.} A single controller is not feasible.
Obviously, a single-point failure may occur and the scalability to expand network is limited because a single controller should ensure the endurance capacity for network flows from various applications and manage a large number of devices.
On the other hand, a distributed control plane can improve the flexibility and resilience of network, e.g., each controller is responsible for a network slice with a certain number of devices.
In the meantime, we also require a distributed control plane to sustain logically centralized network view which is one of important features in SDN.
There are distributed controllers including Onix\cite{koponen2010onix}, HyperFlow\cite{belter2014programmable}, HP VAN SDN[10], ONOS\cite{krishnaswamy2013onos}, DISCO\cite{phemius2014disco}, yanc\cite{monaco2013applying}, PANE\cite{ferguson2013participatory}, and Fleet\cite{matsumoto2014fleet}, but among which few of them can maintain a consistent view of the global network resources.

\textbf{Controller-switch communication authentication.}
The communication channel between controllers and switches suffers from some active attacks in SDN networks, such as man-in-the-middle attack, reply attack and spoofing attack.
Actually, the implementation of OpenFlow originally defined TLS\cite{dierks2008transport} for the controller-switch communication.
However, it made TLS adoption in option on its latter versions due to the high complexity of configuration and high communication cost of TLS\cite{wasserman2013security,son2013model}. This results in lots of security threats such as malicious flow insertion and flow modification when a controller installs a flow to a switch\cite{benton2013openflow}.
Thus, any device connected to a controller under an lightweight authentication protocol is inevitable.
\section{Concrete Design}\label{sec:design}
\begin{figure}[!t]
\centering
\includegraphics[width=5in]{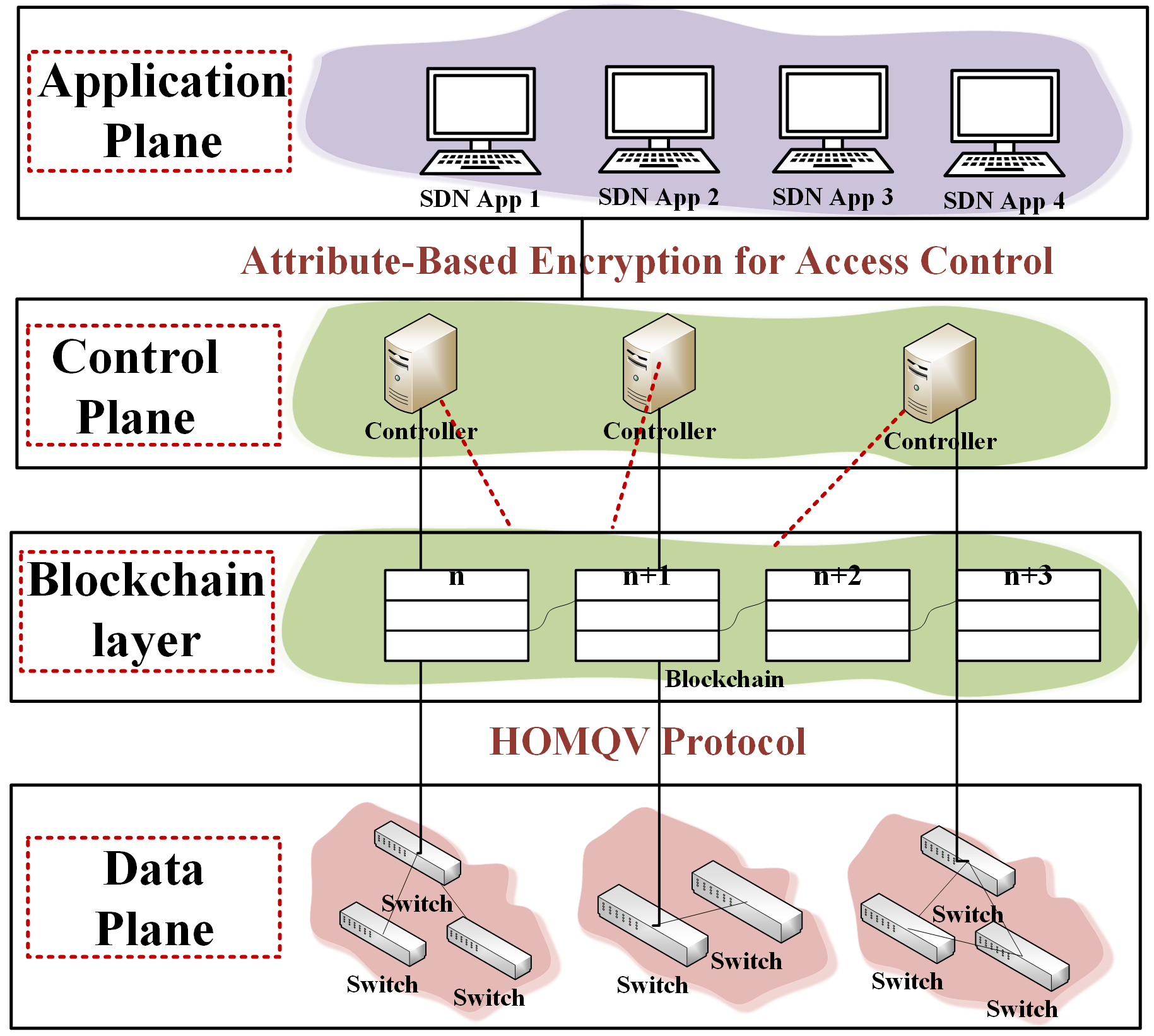}

\caption{The new architecture of SDN appended the Blockchain layer}
\label{fig:overview}
\end{figure}
In this section, we give a concrete design of the Blockchain-based monolithic
secure mechanism as Fig. \ref{fig:overview}, and introduce it from four aspects: Blockchain layer, entities building, transactions building and protocols building.
Additionally, we represent cryptographic primitives an asymmetric encryption algorithm and a digital signature algorithm as $\textsf{AE}$ and $\textsf{DS}$, respectively.
$\textsf{AE}$ algorithm is claimed by $\textsf{(KeyGen, Enc, Dec)}$, the key generation, encryption and decryption algorithms respectively and $\textsf{DS}$ algorithm is defined by $\textsf{(KeyGen, Sig, Ver)}$, the key generation, signature and verification algorithms respectively. A pair of public key and private key in the algorithms are represented as $PK$ and $SK$.
\subsection{Blockchain layer}
Blockchain is used as a packaged and underlying component.
It provides functionalities of resource-recording and resource-sharing among multiple controllers on the control plane.
The functionality of resource-recording represents that Blockchain can be used to record network resources of each controller. The functionality of resource-sharing demonstrates all recorded resources (mainly network events) are shared among all controllers, thereby maintaining the same network view.
We utilize the existing stable Blockchain platform to implement our requirements rather than building a new Blockchain.
The original consensus protocol of the applied Blockchain is not changed and it guarantees the reliability of our new network architecture.
The reason is that many Blockchain-based applications, in order to obtain new required functionalities, create a new Blockchain production (such as utilizing a variant of consensus protocol), but there exist some potential threats, e.g., making chain fork and resulting in a disastrous loss.
Thus, the mechanism applies a worth examining stable Blockchain as the underlying layer of the Blockchain layer.
Blockchain is used to write down all application flows and network events associating with the respective network states where those data are represented as raw transactions.
In addition to transactions, we build smart contract to implement security protocols which further satisfy the security requirements (e.g., to alert the failure of a controller in time). The timestamping and trustworthy features of Blockchain enable the real-time reliability of all recorded application flows and all time-series of the network-wide views during the running process in the arbitrary time.
On the other hand, multiple controllers who are regraded as clients to participate in the underlying Blockchain undertake to record network data (from the application plane and from the device plane) as raw transactions into Blockchain.
The motivation of controllers to manage network, to an extend, keeps the liveness of the underlying Blockchain.
Meanwhile, all applications are obliged to provide network flows (or policies) through the control plane for network devices (e.g., OpenFlow/SDN switches) on the data plane.
OpenFlow/SDN switches also intend to work in network by sending messages (or events) to controllers or providing its resources to controllers.
For example, a switch will request its registered controller when it receives a coming package but fails to forward it. Those network motivations of participating entities enable the underlying Blockchain to be applied significantly.

However, the question on selecting which kind of consensus protocol that the trustworthy underlying Blockchain is based on should be carefully considered.
We desire to gain confirmed transactions (network resources) without canceling on the Blockchain where there does not exist any fork or appear forks with overwhelming probability
and to maintain consistency of those transactions (network resources) among all controllers.
We expect the underlying Blockchain layer to reach consensus under a negligible consensus latency and without the presence of the temporary forks.
The two requirements for the underlying Blockchain imply the property of consensus finality, which is proposed by Vukolic\cite{vukolic2015quest}.
Consensus finality property is defined that once a valid block was appended to the Blockchain at some point in time, the block never was abandoned from the blockchian. \cite{vukolic2015quest} also claimed and proved that any BFT-based Blockchain can satisfy the property of consensus finality, wherein it also supports the excellent network performance and thousands of clients, that is significant to apply to SDN.
\cite{vukolic2016eventually} indicated that there are practical systems (e.g., Ripple network2 \cite{schwartz2014ripple} or OpenBlockchain3 \cite{OpenBlockchain}) implementing the transformation from the eventually consistent POW consensus to the instantly consistent BFT consensus.
Moreover, the transaction processing ability per second is also concerned.
It implies throughput capacity of the Blockchain can perform, which determines the network performance of our module built on the Blockchain (e.g., throughput capacity of the control plane on network events which are network-wide resources).
Fortunately, BFT-based Blockchain protocols enjoy excellent performance on throughput\cite{vukolic2015quest}.
Therefore, a kind of Blockchains based on BFT protocols such as Ripple network2 \cite{schwartz2014ripple} are adopted to implement the Blockchain layer in new architecture of SDN.
\subsection{Entities building}
In our context, we define entities who are actively participating in SDN including applications (i.e., $APP$), controllers (i.e., $CON$) and switches (i.e., $SWITCH$).
Formally, we give expressions to describe the entities in the form of multi-tuples as follows:
\begin{small}
\begin{align*}
  APP & =(\textrm{$ID_{app}$}, \textrm{$PK_{app}$}, \textrm{$SK_{app}$}, \textrm{$category$}) \\
  CON & =(\textrm{$ID_{contr}$}, \textrm{$PK_{contr}$}, \textrm{$SK_{contr}$}, \textrm{$Slice$})\\
  SWITCH & =(\textrm{$ID_{switch}$}, \textrm{$PK_{switch}$}, \textrm{$SK_{switch}$}, \textrm{$Slice$})
\end{align*}
\end{small}Specifically, a unique identifier $ID$ is used to represent the identity of an entity.
An application can use the unique application package name as its identity $ID_{app}$.
For controllers and switches, they use their unique IP addresses or Media Access Control(MAC) addresses, $ID_{contr}$ and $ID_{switch}$ respectively.
A switch works out a pre-setting cryptography puzzle to enable an effective IP address used to register the network.
This method limits the ability of switch to control IP addresses used in the network, which is resistant to spoofing attack.
The tuple $Slice$ represents the network slice a controller or a switch belongs to (The network of SDN composes of several network slices).
Additionally, we employ an asymmetric key generation algorithm, $(\textrm{$PK$}, \textrm{$SK$}) \leftarrow\textsf{KeyGen}$, to create a pair of public key and private key for an entity. According to the autonomous key-selection mechanism used in the context of Blockchain, each entity generates its keys as it desires by the $\textsf{KeyGen}$ algorithm.
The tuple named $category$ in application expression is defined according to use cases among most of SDN applications.

We also define application flows from the application entities and network events created by the switch entities. The entities building flows and creating events meanwhile are responsible to generate identities for them respectively.
\begin{small}
\begin{align*}
  flow & =(\textrm{$ID_{flow}$}, \textrm{$content$}, \textrm{$PK_{app}$}, \textrm{$ID_{contr}$},\textrm{$ID_{switch}$}) \\
  event & =(\textrm{$ID_{event}$}, \textrm{$event$}, \textrm{$PK_{switch}$}, \textrm{$ID_{contr}$}, \textrm{$ID_{switch}$})
\end{align*}
\end{small}The defined tuples for flows and events also indicate where they come from and where they contribute to. When a flow is sent by an App, the App attaches the flow with its signature for the flow \textrm{$Sign_{flow}$} by its private key
\begin{small}
\begin{align*}
  \textrm{$Sign_{flow}$} &= \textrm{$\textsf{DS.Sig}$$(SK_{app},ID_{flow}||ID_{contr}||content)$})
\end{align*}
\end{small}and the same process is necessary for an event.
\subsection{Transactions building}From the time when SDN entities participate in network and during the period they act in network, their all active histories are built into meta-data of transactions on the Blockchain.
The data are classified into three classes: transactions of registered entities, transactions of application flows and transactions of network events.
Transactions of registered entities that not only effectively indicate the existences of entities in SDN but also the connected relationships among them.
At first, each controller managing SDN provides register information which is described in a built entity.
And then they are generated into transactions $T_{contr}$,
\begin{small}
\begin{align*}
  T_{contr} &=(\textrm{$ID_{T_{contr}}$}, \textrm{$ID_{contr}$}, \textrm{$PK_{contr}$}, \textrm{$Slice$}, \textrm{$\textsf{DS.Sig}$$(SK_{contr},$}\\
  &\textrm{$ID_{contr}||Slice)$})
\end{align*}
\end{small} in which $ID_{contr}$ concatenating $Slice$ can be verified by algorithm $\textsf{DS.Ver}$$(PK_{contr}, \textsf{DS.Sig}$$(SK_{contr}, ID_{contr}||Slice))$ that $ID_{contr}$ in the network slice $Slice$ is linked with $PK_{contr}$ and has been registered on the Blockchain layer.
An application connects with a controller to provide network flows for switches which connect with the controller. Thus, the entity information of the application and the relationship information representing that the application connects with the controller will be recorded as $T_{app}$ and $T_{app-contr}$.
\begin{small}
\begin{align*}
  T_{app} &=(\textrm{$ID_{T_{app}}$}, \textrm{$ID_{app}$}, \textrm{$PK_{app}$}, \textrm{$category$}, \textrm{$ID_{contr}$}, \textrm{$Sign_{app}$})\\
  \textrm{$Sign_{app}$} &= \textrm{$\textsf{DS.Sig}$$(SK_{app}$}, \textrm{$ID_{app}||category||ID_{contr})$}\\
  T_{app-contr} &=(\textrm{$ID_{T_{app-contr}}$}, \textrm{$ID_{T_{app}}$}, \textrm{$ID_{T_{contr}}$})
\end{align*}
\end{small}From transactions $T_{app}$, only the legitimate application owning the public key
can succeed to verify $\textsf{DS.Ver}$$(PK_{app}$, $\textsf{DS.Sig}$$(SK_{app},ID_{app}||category||ID_{contr}))$ the respective signature of the identity be accepted by the controller.
Additionally, transactions $T_{switch}$ and $T_{contr-switch}$ are generated once the controller-switch communication is built.
\begin{small}
\begin{align*}
  &T_{switch}=(\textrm{$ID_{T_{switch}}$}, \textrm{$ID_{switch}$}, \textrm{$PK_{switch}$}, \textrm{$Slice$},
  \textrm{$ID_{contr}$}, \textrm{$Com$})\\
  &T_{contr-switch}=(\textrm{$ID_{T_{contr-switch}}$}, \textrm{$ID_{T_{contr}}$}, \textrm{$ID_{T_{switch}}$})
\end{align*}
\end{small}Note that a controller and a switch build an authenticated communication by using HOMQV protocol which has two security problems.
The two problems are that the origin HOMQV protocol fails to against reply attack and the sender of it needs to update a long-term key with a third party
(the long-term key is the key of a switch used to construct a session key with a targeted controller).
We emphasize that the two security issues of the protocol can be overcome with the help of Blockchain.
The transaction $T_{switch}$ contains the information ($ID_{switch}$ and $PK_{switch}$) of a switch who launches a communicated request to a targeted controller following the HOMQV protocol.
Supposed that a compromised switch launches reply attack to a controller by frequently sending its information.
By auditing $T_{switch}$ on the Blockchain, a controller can refuse a replied request when the replied identity ($ID_{switch}$ and $PK_{switch}$) included in the $T_{switch}$.
It is natural to understand because the Blockchain is regarded as a timestamping database and meanwhile stores communication histories among two parties of the protocol.
On the other hand, in the transaction $T_{switch}$, the tuple $Com$
\begin{small}
\begin{align*}
    Com= \textrm{$\textsf{AE.Enc}(PK_{contr}, nonce, \textsf{DS.Sig}(SK_{switch}, nonce))$}
\end{align*}
\end{small} of it is a commitment in encryption by the public key of a controller this switch connects with, which is helpful for the controller to confirm the identity of the switch when its long-term key has been updated.
The commitment includes a nonce the switch selects when the first connection is built.
Specifically, when the same but key-updated switch reconnects with the controller, the controller will audit the transaction $T_{switch}$ on the Blockchain.
It extracts the encrypted tuple from the $T_{switch}$ and makes an identity-verification challenge for the switch. If the switch could answer the correct nonce that the controller verifies whether the encrypted value, with its public key of the nonce is equal to the value extracted from the logged connection transaction. If it does, the verification is effective and the controller continues to share a new session key using its updated key.

An transaction of an application flow, represented by $T_{flow-afore}$ and $T_{flow-after}$, includes the identity of the flow, the flow content, the identifier of the flow-stemming application, the identifier of the targeted controller and a signature signed by the flow-stemming application with its private key. $T_{flow-afore}$ and $T_{flow-after}$ are on behalf of the transactions that some application flow is injected into the network by a specific application, passes by a related controller and ultimately is installed on a specific switch. $T_{flow-afore}$ records the process from some specific application to a specific controller and the other one $T_{flow-after}$ demonstrates its trace from the controller to a specific device.
\begin{small}
\begin{align*}
 T_{flow-afore} &=(\textrm{$ID_{T_{flow-afore}}$}, \textrm{$ID_{flow}$}, \textrm{$ID_{contr}$}, \textrm{$PK_{app}$},
 \textrm{$content$},\\
 &\textrm{$Sign_{flow}$})\\
 \textrm{$Sign_{flow}$} &= \textrm{$\textsf{DS.Sig}$$(SK_{app},ID_{flow}||ID_{contr}||content)$})\\
  T_{flow-after} &=(\textrm{$ID_{T_{flow-after}}$}, \textrm{$ID_{flow}$}, \textrm{$ID_{contr}$}, \textrm{$ID_{switch}$}, \textrm{$state$})\\
  T_{flow} &=(\textrm{$ID_{T_{flow}}$}, \textrm{$ID_{T_{flow-afore}}$}, \textrm{$ID_{T_{flow-after}}$})
\end{align*}
\end{small}Before generating $T_{flow-afore}$, controllers need to authenticate all application flows. A malicious flow will be filtered because controllers will audit the application creating the flow and verify its identity with the transaction $T_{app}$. The transaction $T_{flow-after}$ is recorded until the flow is installed into the flow table of a switch and the transaction contains the respective resource states $state$ sent by the switch where its description is omitted here because this process is similar with the process to generate events by switches as following.

Eventually, the last kind of transactions, that is, transactions of network events dedicate the network events provided by switches when the authenticated communication links have been built. The transactions $T_{event}$ mainly are the dynamic states which are triggered by the respective application events (including in application flows) and additionally \texttt{Pack\_in} messages.
\begin{small}
\begin{align*}
  T_{event} &=(\textrm{$ID_{T_{event}}$}, \textrm{$ID_{event}$}, \textrm{$PK_{switch}$}, \textrm{$ID_{contr}$}, \textrm{$ID_{switch}$}, \textrm{$event$})
\end{align*}
\end{small}Note that an authenticated communication link enables two parties to share a session key and any message among them is protected with the sharing session key. However, messages (network events) included in a generated transaction $T_{event}$ on the Blockchain are in the form of non-encryption. The messages are thought being authenticated because they come from a trustworthy controller-switch communication following the HOMQV protocol.
\subsection{Protocols building}\label{subsec:protocolBuilding}
Relying on the timestamping network records on the Blockchain, we defined necessary protocols to enforce security. In the protocols, a lightweight and efficient ABE scheme and an authentication protocol HOMQV are implemented. With the help of the defined protocols, our security goals are achieved. We explain how to realize the security goals based on the Blockchain and introduce how to combine the Blockchain with ABE scheme and HOMQV protocol.

\textbf{Security enhancement on the control plane} We define protocols of detection and authentication for the newly coming application flows based on the existing transactions on the Blockchain. At first, \texttt{AuthFlowProtocol} is implemented to enable the authentication of application flows. In the protocol a controller first, by $IsRightFlow( )$ verifies the identity of an application creating the flow based on the transactions $T_{app}$, $T_{app-contr}$, $T_{contr}$ and $T_{contr-switch}$. Then, it uses the public key of the application to verify the flow content, by $verifyFlow(PK_{app})$ when the first step is valid.
     \begin{algorithm}[h]
        \caption{\texttt{AuthFlowProtocol}: To check whether the flow are related to the registered application and controller with the records $T_{app}$, $T_{app-contr}$, $T_{contr}$ and $T_{contr-switch}$ on Blockchain. If the relationship records exist, the protocol continues and uses $PK_{app}$ to verify the signature. Then, if the signature is signed by the application and the content of the sent flow does not be modified.}
        \LinesNumbered
        \textbf{procedure}\\
        \textbf{call} \texttt{FlowReplyDetectionProtocol}\;
        \If{ $IsRightFlow()$ $\equiv$ true}
        {
            \If{ $verifyFlow(PK_{app})$ $\equiv$ true}
            {
                 $generateT_{flow-afore}(flow)$\;
            }
            \textbf{end if}\;
        }
        \textbf{end if}\;
        \textbf{end procedure}
     \end{algorithm}
Similarly, \texttt{FlowReplyDetectionProtocol} is used to detect the replied flows by auditing the identifiers of flows based on the transactions $T_{flow-afore}$. It is called by \texttt{AuthFlowProtocol} as its sub-protocol before executing the protocol logic to verify the identity of a flow application. When the flow is installed in a switch by some controller and the associating network states are record, the transaction $T_{flow-after}$ is generated.
     \begin{algorithm}
        \caption{\texttt{FlowReplyDetectionProtocol}: To check $ID_{flow}$ and determine if it has existed. If it has existed, the protocol uses $PK_{app}$ to gain $T_{app}$. If $T_{app}$ exists, the protocol decreases the reputation value of the application as punishment and returns false. If it is a new flow, return true.}
        \LinesNumbered
        \textbf{procedure}\\
        \If{ $checkID_{flow}()$ $\equiv$ true}
        {
            \If{ $getT_{app}(PK_{app})$ $\equiv$ true}
            {
                $reduceReputation(T_{app}.ID_{app})$ \;
                \textbf{return} false\;
            }
            \textbf{end if}\;
        }
        \textbf{end if}\;
        \textbf{return} true\;
        \textbf{end procedure}
     \end{algorithm}
Following the same mechanism as the protocol, the protocol \texttt{AuditNetworkProtocol} provides traceability of network behaviours by auditing the transactions $T_{flow\_afore}$, $T_{flow-after}$  and $T_{event}$.
In particular, it enables the traceability of application flows when it has been injected into network and traces a cause of a network event.
It works because those logged network data (transactions) not only record the sent network flows among SDN entities but also demonstrate all network behaviour which has arose.
Within a running network, if the network suffers abnormal attacks, the attack processes also are logged as transactions.
On the other hand, with those logged transactions of attack trajectories, the future attacks lunched on the network can be recognized that is attacks pattern recognition.
The functionality of the protocol is configured flexibly by network managers according to the need of troubleshooting network.

As for the requirement to support notifications for the applications after a process of flows-arbitration is finished, we define the protocol \texttt{FlowArbitrationLossNotifyProtocol}.
By the protocol, an application lacking arbitration would gain a notification.
It audits the network records that demonstrate which one of conflicted flows targeted at the same switch is adopted. Specifically, it checks the latest record that flow is generated into a transaction $T_{flow-after}$ and $T_{flow-afore}$. After finishing the process to arbitrate conflicted flows, the controller sends a notification to the application generating the arbitrated flow.
    \begin{algorithm}
        \caption{\texttt{FlowArbitrationLossNotifyProtocol}: Send a notification to an application which is out of arbitration.}
        \LinesNumbered
        \textbf{procedure}\\
        $getNewBlock( )$\;
        $ID_{flow}$ $\leftarrow$ $auditT_{flow-after}( )$\;
        $ID_{T_{flow-afore}}$ $\leftarrow$ $auditT_{flow-afore}(ID_{flow})$\;
        $PK_{app}$ $\leftarrow$ $auditT_{flow-afore}(ID_{T_{flow-afore}})$\;
        $ID_{app}$ $\leftarrow$ $auditT_{app}(PK_{app})$\;
        $NotifySDN\_APP(ID_{app})$\;
        \textbf{end procedure}
    \end{algorithm}
In order to ensure the real-time stable response for switches, controllers that the switches link with in SDN need to keep active. Thus,  \texttt{ControllerFailedNotifyProtocol} is defined to notify a switch when its directly connected controller has failed. It depends on an assumption that, if some controller never participates in or becomes off-line, all transactions on the latest several blocks would not demonstrate any network behaviour of the controller. That is, by auditing the transactions of the latest several blocks, the controller are identified being failed to a certain extent. Based on the idea, we define the protocol which is executed automatically once some controller is alive-loss in a period of time on the Blockchain. The protocol needs to periodically check whether all controllers are active by reading records of network behaviours with the latest block being created.
     \begin{algorithm}
        \caption{\texttt{ControllerFailedNotifyProtocol}: It is responsible to send a notification to the switches connecting with a controller when the controller is failed.}
        \LinesNumbered
        \textbf{procedure}\\
        $(T_{event}, T_{flow\_after}) \leftarrow getLastSixBlocks( )$\;
        \If{ $auditAliveOfController$($T_{event}, T_{flow\_after}$) $\equiv$ null}
        {
            \textbf{continue}\;
        }
        \textbf{end if}\;
        \If{ $auditAliveOfController$($T_{event}, T_{flow\_after}$) $\not\equiv$ null}
        {
            $[ID_{T_{contr}}] \leftarrow getFailedControllers( )$\;
            \For{ $ID_{T_{contr}}$ \textbf{in} $[ID_{T_{contr}}]$}
            {
                $[ID_{T_{switch}}]$ $\leftarrow$ $auditT_{contr-switch}(ID_{T_{contr}})$\;
                $[ID_{switch}] \leftarrow getSwitches([ID_{T_{switch}}])$\;
                \For{ $ID_{switch}$ \textbf{in} $[ID_{switch}]$}
                {
                    $NotifySwitch(ID_{switch})$\;
                }
                \textbf{end for}\;
            }
            \textbf{end for}\;
        }
        \textbf{end if}\;
        \textbf{end procedure}
    \end{algorithm}
To stress it once again, we adopt multiple controllers on the control plane while a consistent network view of resources can be maintained using Blockchain as a sharing resource channel. Since all controllers record all network events and collect network resources from devices connecting with it, the network-wide resources are public when the underlying Blockchain announces a new block.
Note that the lastly transactions are viewed valid and accepted consistently under the consensus mechanism of the underlying Blockchain, specifically, the aforementioned BFT-based Blockchain. Therefore, depending on the creation of the reliable new block following the underlying consensus protocol, all controllers achieve consensus on the whole network view.
        \begin{figure}[!t]
        \centering
        \includegraphics[width=5in]{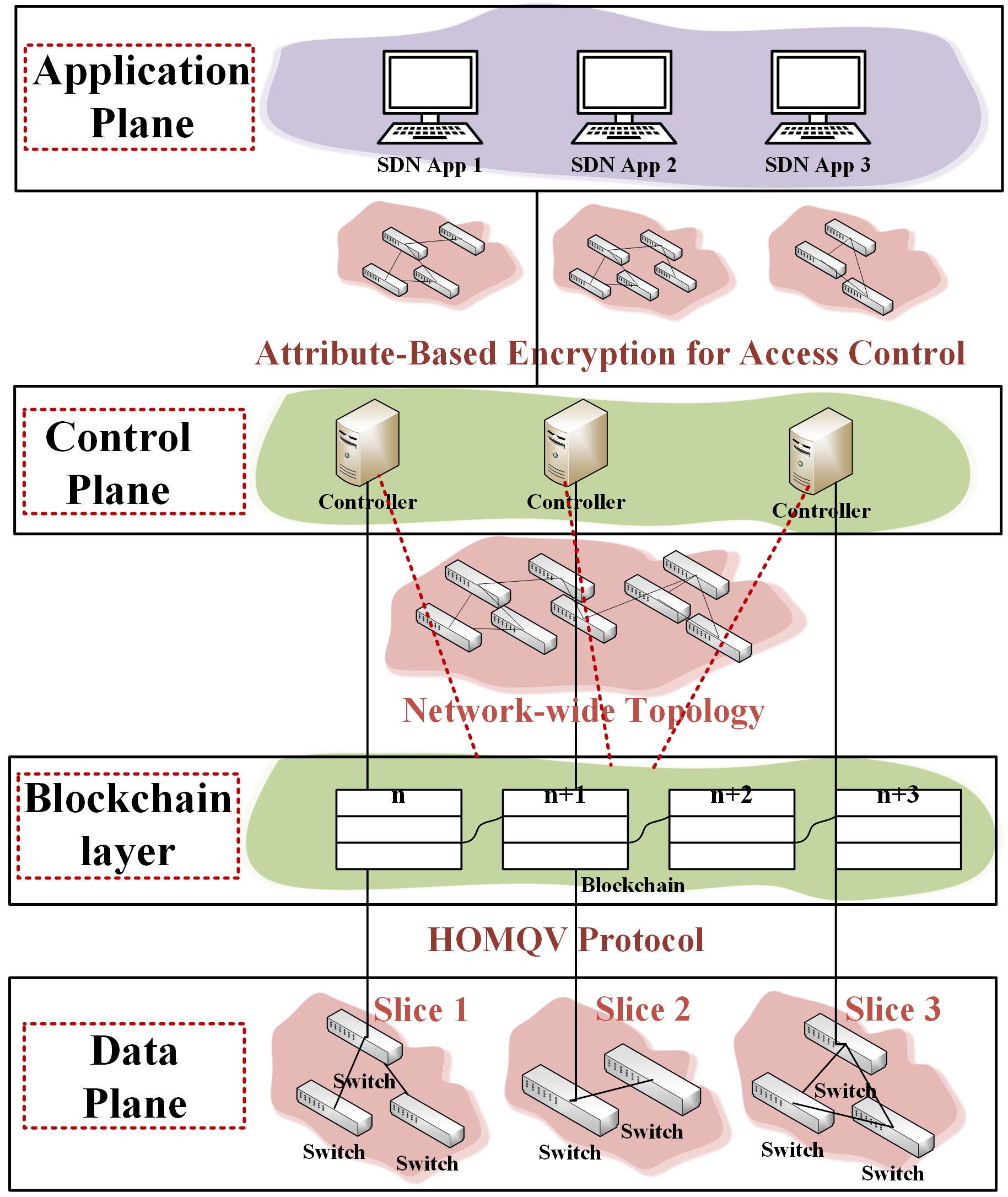}

        \caption{The access control on the network-wide topology resources}
        \label{fig:access}
        \end{figure}

\textbf{Secure assess control on network-wide resources}
\begin{algorithm}
        \caption{\texttt{AccessControlProtocol}: It provides attribute-based access control on network-wide resources with for SDN Apps.}
        \LinesNumbered
        \textbf{procedure}\\
           ${\sf (PK,MK)}$ $\leftarrow$ \textbf{ABE}.${\sf Setup}$()\;
           $getLatestTransactions$( )\;
           $[T_{app}, T_{app-contr}]$ $\leftarrow$ $getTappAndTapp\_contr$( )\;
           \For{ $T_{app}$ \textbf{in} $[T_{app}]$}
            {
                 \For{ $T_{app}$ \textbf{in} $[T_{app-contr}]$}
                {
                    $[T_{contr}]$ $\leftarrow$ $getTapp\_contr$( )\;
                }
                \textbf{end for}\;
                $[Attributes]$ $\leftarrow$ $buildAttrForApp$($T_{app}, [T_{contr}]$)\;
            }
           \textbf{end for}\;
           ${\sf E}$ $\leftarrow$ \textbf{ABE}.${\sf Encryption}$(${\sf TD}$, Attributes,${\sf PK}$)\;
           ${\sf D}$ $\leftarrow$ \textbf{ABE}.${\sf KeyGeneration}$(${\sf AC}$,${\sf PK}$, ${\sf MK}$)\;
           ${\sf M}$ $\leftarrow$ \textbf{ABE}.${\sf Decryption}$(${\sf D}$,${\sf E}$)\;
        \textbf{end procedure}
    \end{algorithm}
    \begin{algorithm}
        \caption{\texttt{AuditAuthenRequestProtocol}: It helps to resist replay attacks of connection requests from a switch.
        If the switch has been connected, it sends three parameters(PK, Cipher, ID\_of\_request);
        otherwise, it sends two parameters(PK, Cipher).}
        \LinesNumbered
        \If{ $checkNumOfParam( )$ $\equiv$ $2$}
        {
            $[T_{switch}]$ $\leftarrow$ $getT_{switch}$( )\;
            $checkResult$ $\leftarrow$ $checkIsReplyRequest$($PK_{switch}$, $C_{AE.Enc}(ID_{switch})$, $[T_{switch}]$)\;
            \If{ $checkResult$ $\equiv$ $false$}
            {
                $T_{switch}$ $\leftarrow$ $buildTransForSwitch$($PK_{switch}$, $C_{AE.Enc}(ID_{switch})$, $ID_{contr}$)\;
                $ID_{T_{switch}}$ $\leftarrow$ $getFromT_{switch}$($T_{switch}$)\;
                $executeHOMQV( )$\;
            }
            \textbf{end if}\;
            \If{ $checkResult$ $\equiv$ $true$}
            {
                $reduceReputation(PK_{switch})$ \;
                \textbf{return} false\;
            }
            \textbf{end if}\;
        }
        \textbf{end if}\;
        \If{ $checkNumOfParam( )$ $\equiv$ $3$}
        {
            \textbf{Call} \texttt{SwitchChallengeProtocol}\;
            \If{ \texttt{SwitchChallengeProtocol} $\equiv$ $true$}
            {
                $executeHOMQV( )$\;
            }
            \textbf{end if}\;
        }
        \textbf{return} true\;
        \textbf{end procedure}
    \end{algorithm}
    \begin{algorithm}
        \caption{\texttt{SwitchChallengeProtocol}: An honest key-updated switch who is intent to rebuild a connection needs to resend a new cipher $Com_{new} = \textsf{AE.Enc}(PK_{contr}, nonce_{new}, \textsf{DS.Sig}(SK_{switch}, nonce_{new}))$ which contains a new nonce and this nonce is required to be equal the nonce which is sent by the switch in the last authenticated connection.}
        \LinesNumbered
        \textbf{procedure}\\
        \textcolor[rgb]{0.50,0.51,0.53}{//get the last tuple of Tswitch, that is a cipher.}\\
        $[T_{switch}]$ $\leftarrow$ $getT_{switch}$( )\;
        $Com$ $\leftarrow$ $T_{switch}.tuple[T_{switch}.length-1]$\;
        $Signature$ $\leftarrow$ $AE.Dec_{SK_{contr}}$($Com$)\;
        $Signature_{new}$ $\leftarrow$ $AE.Dec_{SK_{contr}}$($Com_{new}$)\;
        \If{ $nonce_{new}$ $\equiv$ $nonce$ and
        $\textsf{DS.Ver}(PK_{switch}, Signature))$ $\equiv$ $\textsf{DS.Ver}(PK_{switch}, Signature_{new}))$}
        {
            \textbf{rerun} true\;

            \textbf{end if}\;
        }
        \textbf{end if}\;
        \textbf{return} false\;
        \textbf{end procedure}
    \end{algorithm}We apply ABE scheme to achieve secure access control on the network-wide resources.
Controllers manage the resources with fine-grained access by encrypting each resource with a set of related attributes.
Each individual SDN App keeps a private key associated with an access structure (${\sf AC}$) consisting of a set of attributes and their relations (AND/OR).
Each attribute set is composed of App identities, App functionalities plus the relationships between the Apps and the controllers.
For example, a monitoring App is assigned with an attribute set that includes its identity, its function (i.e., monitoring) and two controller identities it has connected ($Attributes$=$\{ID_{app}, ``monitoring", ID_{contr_{1}}, ID_{contr_{2}}\}$).
In particular, \cite{kreutz2015software} concludes application functionalities of majority of SDN Apps can be classified into the following five functions: traffic engineering; mobility and wireless; measurement and monitoring; security and dependability and data center networking.
After being encrypted, different kinds of network resources are only public to the proper Apps whose access structure related to its private key is satisfied with the set of attributes used to encrypt the kind of resources.
For example, as shown in Fig. \ref{fig:access}, the network resources are network-wide topology diagrams (${\sf TD}$) which are maintained by controllers. By utilizing the fine-grained access control, we encrypt different topology diagrams. Different Apps access the diagrams (topology of some devices rather than all devices) they can decrypt with their private key. Specifically, an example of the access control made on the network resources of topology diagrams is defined by the protocol \texttt{AccessControlProtocol}.
In the example, we take a traffic engineering App ${\sf app1}$ obtaining topology resources of devices as shown in Fig. \ref{fig:access}. We assume ${\sf app1}$ has registered the controller ${\sf contr1}$ and ${\sf contr1}$ manages switches in the ${\sf slice1}$ and ${\sf slice2}$, which means ${\sf app1}$ can access the topology diagrams in the ${\sf slice1}$ and ${\sf slice2}$. Note that the relationships among ${\sf app1}$, ${\sf contr1}$ and switches are recorded as the registered transactions that can be read by ${\sf contr1}$ on the Blockchain. In a word, we need to make an access policy for an App with the network functionality of traffic engineering and undertaking to provide services for switches in the ${\sf slice1}$ and ${\sf slice2}$. On the other hand, an access structure according to a set of attributes owning to ${\sf app1}$ (i.e., $Attribute$ = $\{ID_{app1}, ``traffic\ engineering", ID_{contr1}\}$) is constructed by ${\sf contr1}$. Then, ${\sf contr1}$ executes \textbf{ABE}.${\sf KeyGeneration}$ algorithm with the access structure to generate a private key ${\sf D}$ for ${\sf app1}$. Lastly, ${\sf app1}$ uses ${\sf D}$ to execute \textbf{ABE}.${\sf Decryption}$ algorithm and obtain the topology diagrams.

\textbf{Authenticated controller-switch communication} The authenticated controller-switch communication is implemented by utilizing HOMQV protocol. Meanwhile, in the context of Blockchain, we define two protocols to enhance the aforementioned security issues of HOMQV protocol. On one hand, if being employed directly, HOMQV protocol fails to guarantee a controller resists replay attack launched by a switch who is ready to connect. \texttt{AuditAuthenRequestProtocol} is defined to overcome effectively the issue.
On the other hand, a long term key which is used to key exchange to share a session key, could be self-updated as the switch pleases. At that time, a switch needs to rebuild authenticated communication with the controller it connects last time. The protocol \texttt{SwitchChallengeProtocol} is defined that a key-updated switch proves its existing connection with a controller in SDN and refreshes its connection with the controller.
\section{Security Analysis}\label{sec:analyze}
In this section, we discuss security issues in our mechanism: authentication for application flows, replay attack detection for flows, notification of failed controllers for switches, secure access control for network-wide resources and authentication for controller-switch connection. We first give fivefold secure knowledge which are guaranteed by utilized components in our mechanism.

\textbf{1). The underlying Blockchain layer is health.} Note that our mechanism applies a worth examining stable Blockchain as the underlying layer of the Control plane \cite{schwartz2014ripple, OpenBlockchain}. Thus, that is reasonable for us to believe the underlying Blockchain layer is health, in which its record data are immutable and never abandoned.

\textbf{2). The lightweight ABE scheme is provably secure.} The security of the lightweight ABE scheme is provably secure in the attribute based selective-set model based on the ECDDH assumption, which is demonstrated in the work \cite{yao2015lightweight}.

\textbf{3). The HOMQV protocol is a secure one-pass key-exchange protocol in the random oracle model and under the Gap-Diffie-Hellman (GDH) assumption.} The work \cite{halevi2011one} provides a formal analysis of the protocol's security. Specifically, it assuming the hardness of Diffie-Hellman problem, proves the HOMQV protocol is secure which guarantees sender's forward secrecy and resilience to compromise of ephemeral data.

\textbf{4). The utilized asymmetric encryption algorithm is provably secure.} Our mechanism uses the classic public key cryptosystem \cite{cramer1998practical} presented by Cramer et al. which is provably secure against adaptive chosen ciphertext attack under standard intractability assumptions.

\textbf{5). The utilized digital signature algorithm is \emph{Strongly Existential Unforgeability}.} The public-key signature algorithm such as Schnorr scheme \cite{schnorr1991efficient} satisfies the security notion that an adversary could not output a new message-signature pair ($m^*$, $\sigma^*$) with a totally different $\sigma^*$ even if the adversary has queried signatures on message $m^*$.
\begin{figure}[!t]
        \centering
        \includegraphics[width=5in]{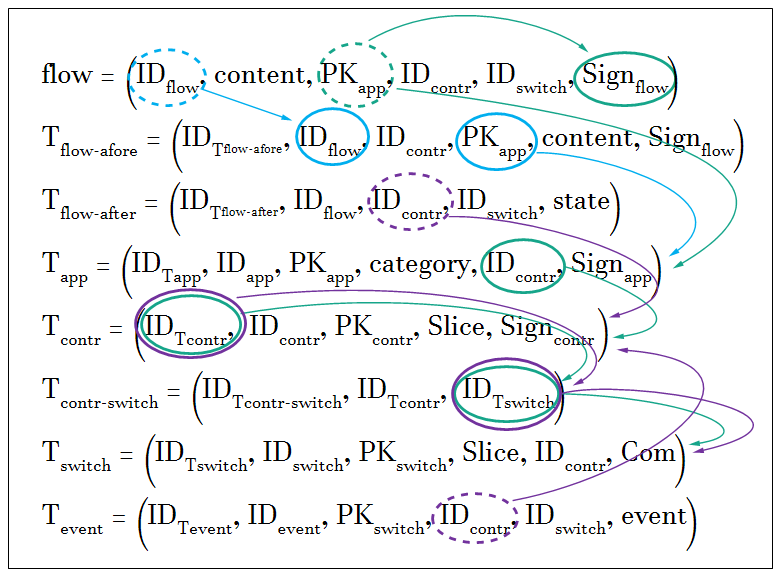}

        \caption{Transaction auditing graph. Note that a circle in dashed line represents a starting point in an auditing process, and directed edges in green lines, blue lines and purple lines are respectively related to the auditing process for authentication for application flows, replay attack detection for flows and notification of failed controllers for switches.}
        \label{fig:auditing}
        \end{figure}

Based on the aforementioned secure knowledge, we present our security analyses with the help of Fig. \ref{fig:auditing}.

\textbf{Authentication for application flows.} The protocol \texttt{AuthFlowProtocol} authenticates the identity of an application flow by identifying whether the flow comes from a legitimate App which has registered the network and  connects with some controller which manages switches in a network slice. A flow (including its identity and the content) is signed by an App with its secret key and then verified by using the public key of the App. The network only accepts legitimate flows but abandons any abnormal flow which is failed to be verified.
$IsRightFlow()$ in the protocol determines whether the
App creating the flow is legitimate based on the transactions $T_{app}$, $T_{app-contr}$, $T_{contr}$ and $T_{contr-switch}$.
Then, it uses the public key of the application to verify the flow signature to determine whether the content flow is modified by
$verifyFlow(PK_{app})$. This process is shown by green circles and green directed edges in Fig. \ref{fig:auditing}. It starts from $PK_{app}$ in the $flow$ and locates the transaction $T_{app}$ this $PK_{app}$ exists. With $ID_{contr}$, it indexes the transaction $T_{contr}$ and then via the relationship transaction $T_{contr-switch}$ the transaction $T_{switch}$ is located. If the process above goes through, it means that the flow comes from the legitimate App and if the transaction in any step of this process does not exist, the flow is rejected. Then, the signature $Sign_{flow}$ is verified by $PK_{app}$ of the App by using the verification algorithm of digital signature algorithm $\textsf{DS.Ver}$$(PK_{app}, Sign_{flow})$.

\textbf{Replay attack detection for application flows.}
The protocol \texttt{FlowReplyDetectionProtocol} detects replayed flows based on the logged records $T_{flow-afore}$ on the Blockchain. When a newly coming flow is received, the identity of the flow is detected by auditing the transaction $T_{flow-afore}$ as shown in blue circles and blue directed edge.
The flow is accepted if the flow has never been sent. Otherwise, the flow is rejected and the App sending this flow is punished by locating the transaction $T_{app}$.

\textbf{Notification of failed controllers for switches.}
The protocol \texttt{ControllerFailedNotifyProtocol} can notify the switches when the controllers being connected break down. By auditing the records of network behaviors $T_{flow-after}$ and $T_{event}$ within the latest 6 blocks, controllers $ID_{contr}$ without any active response can be found out. Then, based on the relationship transaction $T_{contr-swith}$, switches $ID_{switch}$ connecting with the failed controllers would be notified as shown in purple circles and purple directed edges.

On the other hand, the rest of two security issues are analyzed as follows.

\textbf{Secure access control on network-wide resources.}
The protocol \texttt{AccessControlProtocol} implemented by ABE scheme in  \cite{yao2015lightweight} enables an App to access the respective resources when the attribute set of the App satisfies the access structure related to the encrypted resources. Based on the acquired security knowledge that this ABE is provably secure, the access control mechanism is secure.

\textbf{Authentication for controller-switch connection.}
The protocol \texttt{AuditAuthenRequestProtocol} based on HOMQV protocol which has been proved security implements the authenticated communication between controllers and switches. With an extra protocol \texttt{SwitchChallengeProtocol}, two existing security issues of the origin HOMQV protocol are worked out. The two protocols implemented on the Blockchain provide secure authentication enhancement of controllers and switches.
\section{Proof-of-concept Implementation}\label{sec:implement}
       \begin{figure}[!t]
        \centering
        \includegraphics[width=5in]{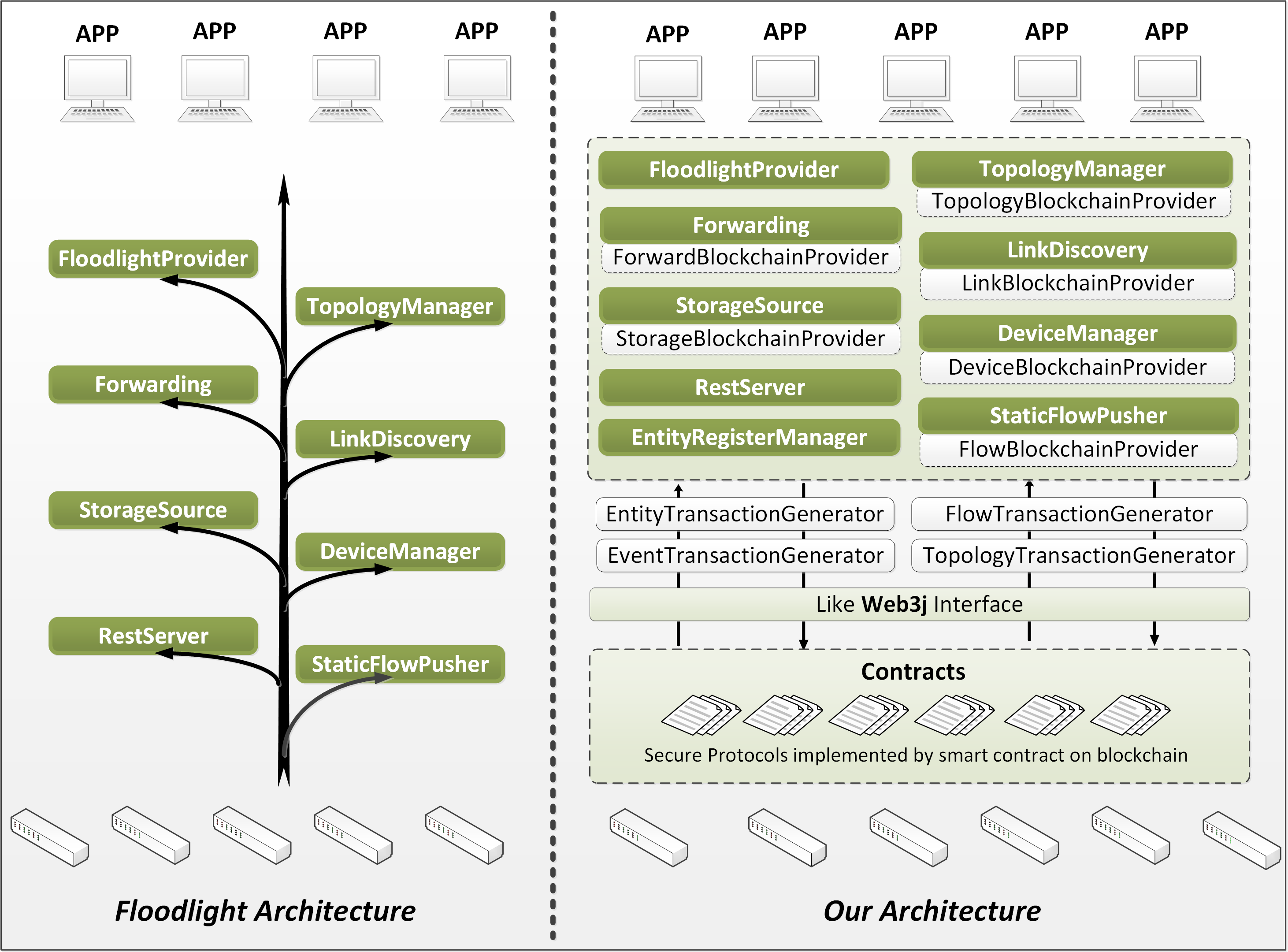}

        \caption{Schematic of our architecture prototype}
        \label{fig:topology}
        \end{figure}
As Floodlight \cite{floodlightcontroller} project puts the world's largest SDN ecosystem into practice\cite{Ecosystem}, we decide to build our mechanism on the Floodlight project and illustrate the utility of our security enhancement.
On the other hand, we build Blockchain environment based on Hyperledger Fabric $V1.0$\cite{Hyperledger} which is an open project of Blockchain. As shown in Fig. \ref{fig:topology}, we demonstrate a schematic of our architecture prototype and compare it with original Floodlight architecture.
Focusing on the security goal we intend to achieve, we append our required Blockchain providers to the corresponding Floodlight application modules, in which Blockchain providers are implemented surrounding Floodlight application modules by applying programming method of Aspect Oriented Programming.
The Blockchain providers, \texttt{TopologyBlockchainProvider} and \texttt{LinkBlockchainProvider} are attached to the primary modules \texttt{TopologyManager} and \texttt{LinkDiscovery} respectively. \texttt{TopologyBlockchainProvider} collects topology resources of network via \texttt{TopologyManager} while \texttt{LinkBlockchainProvider} monitors the status of links in network.
In the meantime, they are responsible to communicate with the defined security protocol \texttt{AccessControlProtocol} on Blockchain so that a customized access control mechanism for applications is provided.
The provider named \texttt{ForwardBlockchainProvider} undertakes to monitor network packages which are forwarded among devices. It collects package information and forwarding pathes that are prepare for flow transaction and event transaction generation.
\texttt{DeviceBlockchainProvider} tracks information of network devices, which is used to generate entity transactions.
\texttt{FlowBlockchainProvider} is used to catch the newly flows via \texttt{StaticFlowPusher}, which is prepare for \texttt{FlowTransactionGenerator}.
Between the communication of module-appended Floodlight project and Blockchain, we construct an implemented project decoupling from the Floodlight project which includes four modules. The four modules are \texttt{EntityTransactionGenerator}, \texttt{FlowTransactionGenerator}, \texttt{EventTransactionGenerator} and \texttt{TopologyTransactionGenerator} which undertake to encapsulate network data and generate transactions in the context of Blockchain.
In addition, the security protocols mentioned in section \ref{subsec:protocolBuilding} are implemented by smart contract which is validated and secure on Blockchain.
In order to connect with Blockchain, this exists an interface like \texttt{Web3j} devoted to build defined transactions and security protocols into Blockchain.
Note that Ethereum\cite{Ethereum} which is another open source Blockchain project offers  a library called \texttt{Web3j} for a variety of Jave application integrating with Ethereum. For Hyperledger Blockchain, the third-party library, as \texttt{Web3j} for Ethereum, also is expected to be used to integrate our Java application based on Floodlight. By calling the third-party library as middle interface between Floodlight project and Blockchain project, the communication with secure protocols on Blockchain can be achieved.
\section{Conclusion}\label{sec:conclusion}
SDN has become an emerging technology to enhance network performance. With its extensive adoption, some security issues of SDN are exposed and imperative to be studied. In the paper, we present a Blockchain-based monolithic secure mechanism for SDN. By utilizing Blockchain to record all network flows and events and to implement secure protocols with smart contracts, the presented secure mechanism overcomes the common security issues in SDN.  In particular, the decentralized control plane tackles the problem of single-point failure and improves network scalability; application flows can be authenticated, tracked and accounted; network-wide resources are protected with access control scheme and authenticated communication channels are ensured between controllers and switches. At last, the security analysis and an implementation prototype of our mechanism demonstrate the effectiveness of security improvement for SDN.

\bibliographystyle{unsrt}
\bibliography{foo}  

\end{document}